\documentclass{article}

   \PassOptionsToPackage{numbers, sort, comma, square, compress}{natbib}

     \usepackage[preprint]{neurips_2020}

\usepackage[utf8]{inputenc} 
\usepackage[T1]{fontenc}    
\usepackage{hyperref}       
\usepackage{url}            
\usepackage{booktabs}       
\usepackage{amsfonts}       
\usepackage{nicefrac}       
\usepackage{microtype}      
\usepackage{microtype}
\usepackage{graphicx}
\usepackage{booktabs} 
\usepackage{amsfonts} 
\usepackage{nicefrac} 
\usepackage{microtype} 
\usepackage{array}
\usepackage{longtable}
\usepackage{amsthm}
\usepackage{multirow}
\usepackage{float}
\usepackage{amsfonts}
\usepackage{hyperref}
\usepackage{csvsimple}
\usepackage{mathtools}
\usepackage{tikz}
\usepackage[]{algorithm2e}
\usepackage{tabularx}
\usepackage{rotating}
\usepackage{ulem}
\usepackage{color}
\usepackage{subfig}
\usepackage{amssymb}
\usepackage{dsfont}
\usepackage{theoremref}

\newcommand*{\Z}{\mathds{Z}}
\usepackage[T1]{fontenc}
\newcommand*{\bX}{{\bf X}}
\usepackage{mathtools,bm,amsfonts,amssymb,lmodern}

\usepackage{marvosym}

\usepackage{cleveref}
\usepackage[outercaption]{sidecap}    
\usepackage{hyperref}

\title{Causal analysis of Covid-19 spread in Germany}

\author{%
    Atalanti A. Mastakouri \\
  Department of Empirical Inference\\
  Max Planck Institute for Intelligent Systems\\
  T{\"u}bingen, Germany\\
  \texttt{atalanti.mastakouri@tuebingen.mpg.de}
   \And
  Bernhard Sch{\"o}lkopf \\
  Department of Empirical Inference\\
  Max Planck Institute for Intelligent Systems\\
  T{\"u}bingen, Germany\\
  \texttt{bs@tuebingen.mpg.de}
}
\makeatletter
\newcommand{\mylabel}[2]{#2\def\@currentlabel{#2}\label{#1}}

\newcommand{\xdashleftrightarrow}[2][]{\ext@arrow 3359\leftrightarrowfill@@{#1}{#2}}
\def\leftrightarrowfill@@{\arrowfill@@\leftarrow\relbar\rightarrow}
\def\arrowfill@@#1#2#3#4{%
 $\m@th\thickmuskip0mu\medmuskip\thickmuskip\thinmuskip\thickmuskip
 \relax#4#1
 \xleaders\hbox{$#4#2$}\hfill
 #3$%
}
\renewcommand{\@algocf@capt@plain}{above}
\makeatother
\newtheorem{theorem}{Theorem}

\crefname{lemma}{Lemma}{Lemmas}
\newtheorem*{definition}{Definition}

\makeatother
\begin{document}

\maketitle

\newcommand{\dd}[1]{\mathrm{d}#1}
\begin{abstract}
In this work, we study the causal relations among German regions in terms of the spread of Covid-19 since the beginning of the pandemic, taking into account the restriction policies that were applied by the different federal states. We propose and prove a new theorem for a causal feature selection method for time series data, robust to latent confounders, which we subsequently apply on Covid-19 case numbers. We present findings about the spread of the virus in Germany and the causal impact of restriction measures, discussing the role of various policies in containing the spread. Since our results are based on rather limited target time series (only the numbers of reported cases), care should be exercised in interpreting them. However, it is encouraging that already such limited data seems to contain causal signals. This suggests that as more data becomes available, our causal approach may contribute towards meaningful causal analysis of political interventions on the development of Covid-19, and thus also towards the development of rational and data-driven methodologies for choosing interventions. 
\end{abstract}

\section{Introduction}
The ongoing outbreak of the Covid-19 pandemic has rendered the tracking of the virus spread a problem of major importance, in order to better understand the role of the demographics and of political measures taken to contain the virus.  Until 15/5/2020, 175,699 cases and 8,001 deaths were recorded in Germany, a country with a population of 83 million people, 16 federal states with independent local governments, and 412 districts (Landkreise).  In this paper, we focus on a causal time series analysis of the Covid-19 spread in Germany, aiming to understand the spatial spread and the causal role of the applied restriction measures. 

Causal inference from time series is a fundamental problem in data science, and many papers provide solutions for parts of the problem subject to necessary assumptions \cite{runge2019detecting,pfister2019invariant,eichler2007causal,malinsky18a,Entner2010OnCD,mastakouri2020necessary}. The main difficulty in this research problem is the possibility of hidden confounding in the data, as it is almost impossible in real datasets to have observed all the necessary information. Another problem is a characteristic of the time series themselves, which create dependencies due to connections in the past, hindering the formulation of necessary d-separation statements for graphical inference \cite{pearl2009}. Finally, many known methods cannot handle instantaneous effects that may exist among the time series.

We consider a problem with small sample size compared to the dimension of its covariates, yet of significant current importance: the tracking of the spread of the Covid-19 pandemic, based only on the reported cases. Not having access to all relevant covariates and to all interventions that were applied at different times by different regions constitutes a heavily confounded problem, whose causal analysis requires a method which is robust to hidden confounders. Tracking the Covid-19 spread is of interest since it may help understand and contain the virus. There are significant efforts to understand this based on individual location or proximity information \cite{scantamburlo2020covid}. Other efforts try to understand and quantify the importance of applied restriction measures through modelling of the spread \cite{Dehningeabb9789,kraemer2020effect,chinazzi2020effect,fanelli2020analysis}. In the present work, we focus on the causal analysis of the spread. We perform an offline causal inference analysis of the reported daily Covid-19 case numbers in regions of Germany, in combination with the restriction measures that were applied to contain the spread. 

The most common  and established approach for causal inference on time-series is Granger causality \cite{Wiener1956, Granger1969, Granger1980}. In the multivariate case, we say that $X^j$ {\it Granger-causes} $X^k$ $(k\ne j)$ if a conditional dependence $X^k_t \not \!\perp\!\!\!\perp X^j_{\text{past}(t)} \mid \bX^{-j}_{\text{past}(t)}$ exists (here, $-j$ denotes all indices other than $j$, and past$(t)$ denotes all indices $t'<t$). The fundamental disadvantage of this method is its reliance on \textit{causal sufficiency}: the assumption that all the common causes in the system are observed; in other words, that no hidden confounders can exist \cite{Spirtes1993}. Violations of this, common in real world data, render Granger causality and its extensions \citep[e.g.][]{hung2014,GUO200879} incorrect, yielding misleading conclusions.

Below, we propose and prove a theoretical extension that relaxes the stricter assumption of the SyPI algorithm \cite{mastakouri2020necessary}, a causal feature selection method for time series with latent confounders. We apply it on Covid-19 cases reported by German regions, with the goal to detect which regions and which restriction policies played a causal role on the formation and modulation of the regionally-reported daily cases. We perform this analysis on a state and on a district level. We compare our findings with predictions of the widely used Lasso-Granger method \cite{Arnold2007}, showing that SyPI yields more meaningful results.
Note that while no ground truth exists, our detected causes tend to be neighbouring states/regions, with discrepancies that can often plausibly be attributed to the existence of major transportation hubs.

\section{Methods and Tasks}
\subsection{Causal inference on time series}
\begin{figure}[H]
\centering
\includegraphics[width=0.5\columnwidth]{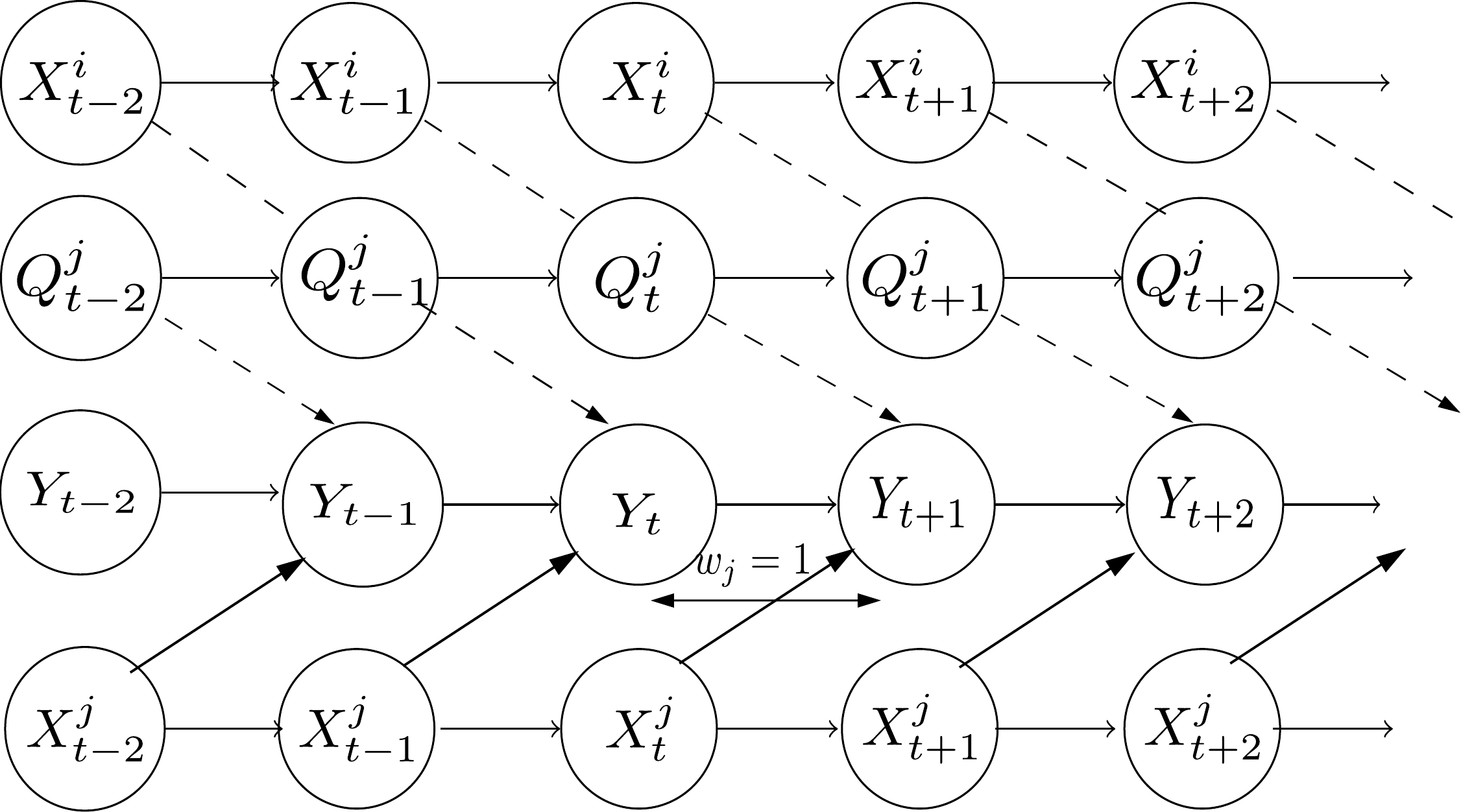}
\caption{An example full time graph of two observed ($X^i, X^j$), one potentially hidden ($Q^j$)  and one target ($Y$) time series.}\label{summarygraph}
\end{figure}
For the problem of causal feature selection on time series data, we are given observations from a target time series $Y:= (Y_t)_{t\in \Z}$ whose causes we wish to find, and observations from a multivariate time series $\bX:=((X^1_t,\dots,X^d_t))_{t\in \Z}$ of potential causes (candidate time series). In settings were Causal sufficiency cannot be assumed, like the one we tackle here, unobserved multivariate time series, which may act as common causes of the observed ones also exist. An example of such a setting is given in Figure \ref{summarygraph}.

\subsection{The SyPI method}
Here we use and theoretically extend the SyPI method proposed by \cite{mastakouri2020necessary}, as it can give causal conclusions in large, dense graphs of time series, based solely on observational data and without assuming causal sufficiency.
According to \cite{mastakouri2020necessary}, the method requires, as input, observations from a target time series $Y$  and from a multivariate time series (candidate causes) $\bX$, as defined above. Moreover, it allows for unobserved multivariate time series, which may act as hidden confounders. Under suitable assumptions discussed in Section~\ref{assumptions}, the method provably detects all the direct causes of the target and some indirect ones, but never confounded ones (its conditions are both necessary and sufficient\footnote{Although SyPI's conditions are necessary only for single-lag dependencies, the method has provided satisfying results even with multiple lags \cite{mastakouri2020necessary}. The existence of multiple lags would only result in fewer detected causes, without affecting the validity of the method in terms of false positives. 
}). 

We now try to provide some intuition. Since it requires familiarity with (and terminology of) causal structure learning, some readers may want to consult \cite{mastakouri2020necessary}.
For each candidate causal time series $X^i$ that has a dependency with the target $Y$ at lag $w_i$, the method performs targeted isolation of the path $X^i_{t-1} \rightarrow X^i_{t}$ - -$Q^j_{t'} \dashrightarrow Y_{t+w_i}$ (where $w_i \in Z, t'<t+w_i, Q^j \in \mathbf{X^{-i}}$ or unobserved), that contains, for every candidate $i$, the current $X^i_{t}$ and the previous time step $X^i_{t-1}$ of the candidate causal time series, and the corresponding node of the target time series $Y_{t+w_i}$.\footnote{'$\dashrightarrow$' denotes a directed path, '- -' denotes a collider-free path.} It does so by building a conditioning set that contains the nodes of $\mathbf{X^{-i}}$ that enter node $Y_{t+w_i-1}$ (temporal ancestor of the target node $Y_{t+w_i}$ of the same time series), including the node itself. This way, it exploits the fact that if there is a confounding path between $X^i_{t}$ and $Y_{t+w_i}$, then $X^i_{t}$ will be a collider that will unblock the path between $X^i_{t-1}$ and $Y_{t+w_i}$ when we condition on it. Therefore, running SyPI boils down to testing two conditions: condition 1 examines if $X^i_{t}$ and $Y_{t+w_i}$ are conditionally dependent given the aforementioned conditioning set, and condition 2 examines if $X^i_{t-1}$ and $Y_{t+w_i}$ become conditionally independent if $X^i_t$ is included in the aforementioned conditioning set. If both conditions hold true, then SyPI identifies $X^i$ as a cause of $Y$ \cite{mastakouri2020necessary}.

\subsection{Weakening SyPI's assumptions}\label{assumptions}
According to \cite{mastakouri2020necessary} SyPI is a sound and complete causal feature selection method in the presence of latent common causes subject to certain graph restrictions. Among the most important graphical assumptions required is that the target be a sink node (assumption 6 in \cite{mastakouri2020necessary}), i.e., the target has no descendants. In Theorem \ref{theorem3} below, we relax this strict assumption, proving that it suffices that none of the (direct or indirect) descendants of the target belongs in the pool of the candidate causes. While this relaxation is important for our application, we prefer not to repeat all assumptions and definitions from \cite{mastakouri2020necessary}. Rather, we describe below what needs to be adapted to handle our more general setting. 

The intuition behind Theorem \ref{theorem3} is the following. The original assumption 6 ensures that when an unconfounded path $X^i_t \rightarrow Y_{t+w_i}$ for some lag $w_i$ exists, the true cause $X^i$ will not be rejected due to a parallel path $X^i_t \rightarrow X^j_{t'} \leftarrow Y_{t+w_i}$ that contains a collider $X^j_{t'}$, which could potentially be unblocked rendering condition 2 of Theorem 2 in \cite{mastakouri2020necessary} false. Theorem 1 of \cite{mastakouri2020necessary} remains unaffected from whether $Y$ is a sink node or not, because in the case that it is not, i.e., $X^i_t \leftarrow Y_{t+w_i}$, condition 2 will correctly reject $X^i_t$. 

Therefore, we only need to show that Theorem 2 of \cite{mastakouri2020necessary} remains unaffected if instead of $Y$ being a sink node, all of its descendants do not belong in $\mathbf{X}$ (we write $\mathbf{DE}_Y^{\mathcal{G}} \not \in \mathbf{X}$). In the case that all the descendants of $Y$ do not belong in its candidate causes $\mathbf{X}$, then they will be unobserved. Assume there is one descendant $D\not \in \mathbf{X}$ of $Y$ that is also connected with a node $X^i_t$ from $\mathbf{X}$. Then $D$ can only have incoming arrows from $X^i_t$ and therefore $D$ is an unobserved collider (any out-coming arrow from $D$ to $\mathbf{X}$ will violate the assumption $\mathbf{DE}_Y^{\mathcal{G}} \not \in \mathbf{X}$). Therefore any path that contains the unobserved collider $D$ cannot be unblocked to create any additional dependencies, because $D$ and any of its descendants cannot belong in the conditioning set. 

\begin{theorem}[Theorems 1 and 2 from \cite{mastakouri2020necessary} still apply]\label{theorem3}
Given the target time series $Y$ and the candidate causes $\mathbf{X}$, assuming Causal Markov condition, causal faithfulness, no backward arrows in time $X^i_{t'} \not \rightarrow X^j_t, \forall t'>t, \forall i,j$, stationarity of the full time graph as well as assumption A7-A9 from \cite{mastakouri2020necessary}, if the target $Y$ is not a sink node, but, instead, none of its direct or indirect descendants belongs in $\mathbf{X}$: $\mathbf{DE}_Y^{\mathcal{G}} \not \in \mathbf{X}$, then Theorem 1 and 2 from \cite{mastakouri2020necessary} still apply. That means the conditions of Theorem 1 from \cite{mastakouri2020necessary} are still sufficient for identifying direct and indirect causes, and conditions of Theorem 2 from \cite{mastakouri2020necessary} are still necessary for identifying all the direct unconfounded causes. 
\end{theorem}
We prove Theorem \ref{theorem3} in the Appendix (Section \ref{proof}).

While the relaxed assumption makes the result more generally applicable, we need one additional step to apply it to our dataset: The algorithm requires as input the candidates and the target as two separate variables. Therefore, we need to assign one region at a time as \textbf{target}. In order to comply with the aforementioned assumption, instead of directly feeding all the remaining time series as the candidate causes of the target $Y$, we use as \textbf{candidate causes} those other regions that \textbf{have reported Covid-19 cases before the target} (in addition to the applied policies for the analysis at the federal states level). This makes it more likely that no effects of the target exist in its candidate causes (assuming stationarity of the graph). 

SyPI assumes that the causal relations among the time series are stationary, not changing in different time windows. However, since we do not know the ground truth, it is possible that the policies not only cause the reported infections time series but also be caused by it in different time windows. This possible violation of stationarity of the graph creates problems because it also implies arrows from the target to some of the policies time series which belong in its candidate causes. Therefore this could violate both the assumption 6 in \cite{mastakouri2020necessary} about the target being a sink node and the relaxed proposed assumption $\mathbf{DE}_Y^{\mathcal{G}} \not \in \mathbf{X}$. We are aware that this could happen, which is one reason we are careful in our conclusions. 

The source code for the analysis presented here can be found in the supplement.
\subsection{Selection of statistical thresholds}
Since the causal Markov condition and causal faithfulness are assumed (definitions \ref{defCMC}, \ref{faithfulness}), there is an equivalence between d-separation statements in the graph and conditional independences on the probability distributions of the variables. As \cite{mastakouri2020necessary}, we use SyPI for linear relationships only (although the theory is more general), and hence resort to partial correlations to test the conditional dependence (condition 1) and the conditional independence (condition 2) of \cite{mastakouri2020necessary}.
SyPI operates with two thresholds for those two tests: one for rejecting conditional independencies (condition 1), and another for accepting conditional independencies (condition 2). Since the time series of the daily reported cases since the beginning of the pandemic in Germany include only 87 reported days, we decided to explore the outcomes of the algorithm for stricter and looser thresholds. We thus examined values of threshold-1 in $\{0.01, 0.05\}$ and values for threshold-2 in $\{0.1, 0.2\}$. We report the causal findings for the looser combination $(0.05, 0.1)$ in Fig.~\ref{foundedCausesWithPolicies} and for all four in the supplement (Fig.~\ref{federal_policies_all}).

\section{Experiments}
\subsection{Dataset: Daily reported Covid-19 cases for German regions}
\begin{figure}[H]
\begin{center}
\includegraphics[width=0.95\linewidth]{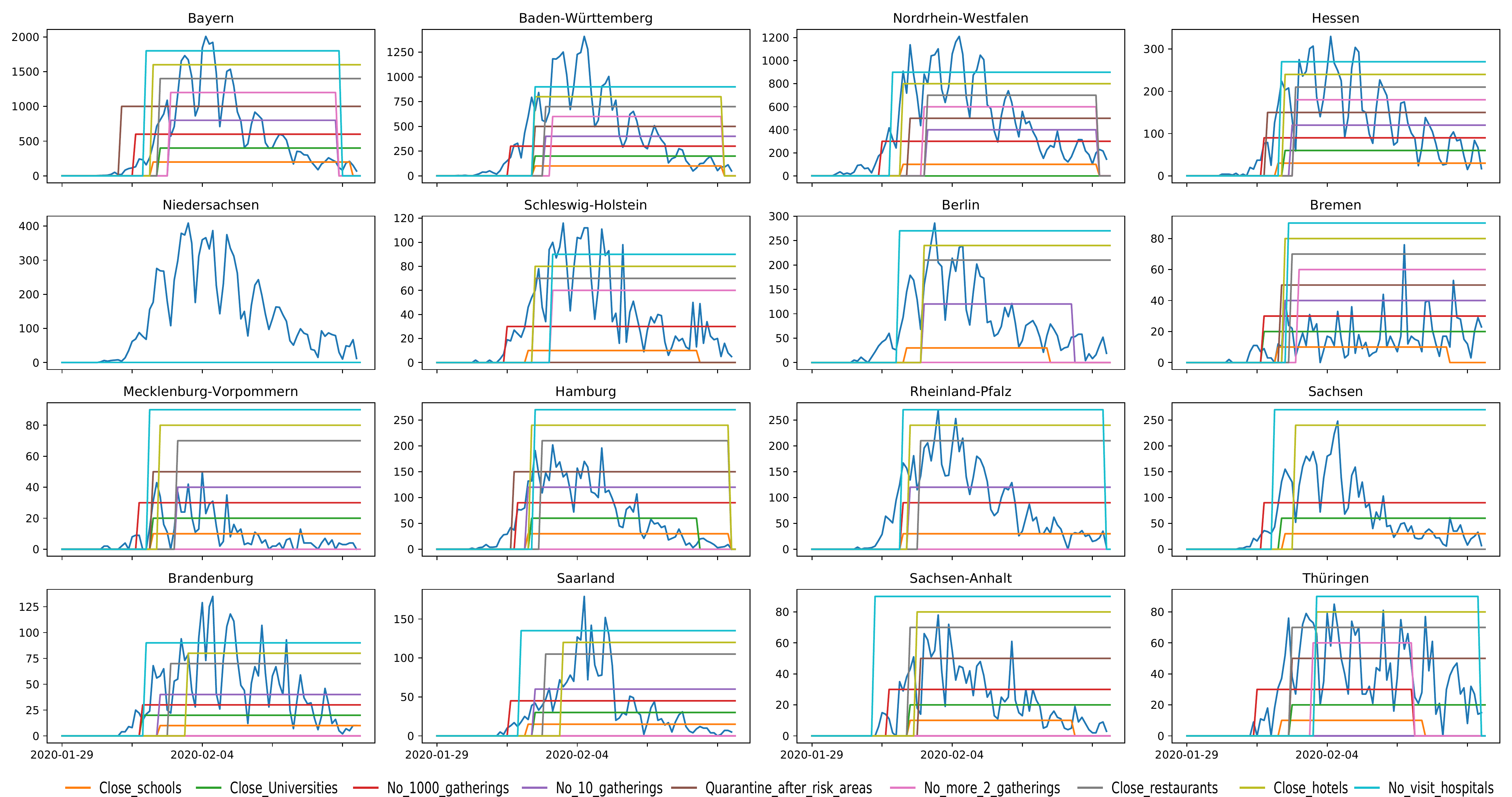}
\caption{Time series of daily reported detected Covid-19 cases in each federal state from 28/01/2020 until 15/05/2020. The blue curve represents the daily reported infections as a function of time. In addition to the Covid-19 cases, 9 restriction measures are depicted as indicator functions (see legend and main text). The height of the indicator functions does not have a meaning. It is only adjusted for visibility purposes.}\label{fallsandpolicies}
\end{center}
\end{figure}

The data are taken from the official reports of the Robert-Koch Institute, last downloaded on 15/05/2020 \cite{RKIdataset}. They are analysed in two steps:
\paragraph{Causal analysis on federal state level}\label{federal}
Figure \ref{fallsandpolicies} depicts daily reported Covid-19 cases for each of the 16 German federal states, each one represented by a time series, starting from when the first report was made (28/01/2020) until 15/05/2020. The plots are sorted chronologically, with the top left corresponding to the Bundesland (federal state) that reported first, and the bottom right the Bundesland that reported Covid-19 cases last. In addition, we created indicator functions for nine restriction measures that were imposed separately in each Bundesland, as gathered from the official German states' websites and from \url{https://calc.systemli.org/u0o26ims15cr}. The periods these measures were in effect are depicted as indicator functions (vertically scaled to make sure all are visible) in the above plots. The policies are: \textit{closing of schools, closing of universities, ban of gatherings of more than 1000 people, ban of gatherings of more than 10 people, obligatory quarantine of 14 days after returning from risk areas, ban of gatherings of more than 2 people, closing restaurants, closing hotels, forbidding visits in hospitals and nursing homes}. We provide the data in the supplement and here \url{https://owncloud.tuebingen.mpg.de/index.php/s/r4dPdpSBAzP6Ee5}. Note that not all policies were applied in all federal states, and also that for the state of Niedersachsen, no policies are provided. We apply the algorithm for each target state independently, keeping as candidates all the federal states that have reported cases before the target one, as well as the nine aforementioned policies for the specific target. Results are shown in Figure \ref{foundedCausesWithPolicies}.

\paragraph{Causal analysis on district level}\label{district}
To get further results, we apply the modified SyPI method on the time series of daily reported Covid-19 cases for all 412 districts of Germany. We apply it the following way: For every district we use SyPI twice; the first time using as candidate causes all the neighbouring districts of the target that have reported cases before, and the second time, using the same number of districts but from random non-neighbour (distant) locations that have also reported cases before the target. Our hope would be that SyPI will identify more causes among the neighbour districts than among the non-neighbour ones. Furthermore, we would hope that (some of) the latter could be justified by a large airport close-by. The default thresholds of SyPI $(0.01, 0.2)$ were used. For this analysis, we created a matrix with all the neighbour districts of each district, as well as the location of the largest airports (including the number of flights from the past years), which we also provide in the supplement. Furthermore, for the largest airports in terms of number of passengers per year, according to the German flight security organisation (DFS) (MUC, STR, TXL, FDH, FMM, NUE, HAM, FRA, HHN, HAJ, NRN, CGN, DUC, DMT, DRS, BRE, KSF, SCN), we check which districts are near (within 40km) each one of these. In total, 169 out of the total 412 districts were found to be near one of the large airports. The 40km distance was chosen as it corresponds to the diameter of a medium size German district. We then categorise our results in four categories: 1. Detected causes among the neighbours of the target, 2. Detected causes near (within 40km) the target, 3. Detected causes near (within 40km) a large airport, 4. Distant targets that cannot be categorised otherwise.

\subsection{Comparison against Lasso-Granger}
As a baseline for the modified SyPI method for the spread of Covid-19 in the German federal states, we use Lasso-Granger \citep{Arnold2007}. Granger causality is the most widely used method for causal time series analysis, although it assumes causal sufficiency, which we expect to be heavily violated in real data.

\section{Results}
\subsection{SyPI on Covid-19 cases and policies in the federal states}
In Figure \ref{foundedCausesWithPolicies}, the policies and federal states that were identified as causes by SyPI are depicted with target/color specific arrows. Fig. \ref{foundedCausesWithPolicies} correspond to the ``looser'' combination of thresholds $(0.05, 0.1)$. Figure \ref{federal_policies_all} (supplement) provides results for all four combinations. As we can see, the result does not change dramatically with different threshold combinations, but as expected, more causes are detected with the ``looser'' combination $(0.05, 0.1)$, as it more easily accepts dependencies and independencies. We discuss the findings in Section~\ref{results_discussion}.

\subsection{SyPI vs Granger}
\begin{figure}[H]
  \centering
  \subfloat[Federal state level \& policies causal analysis.\label{foundedCausesWithPolicies}]{{\includegraphics[width=.5\textwidth]{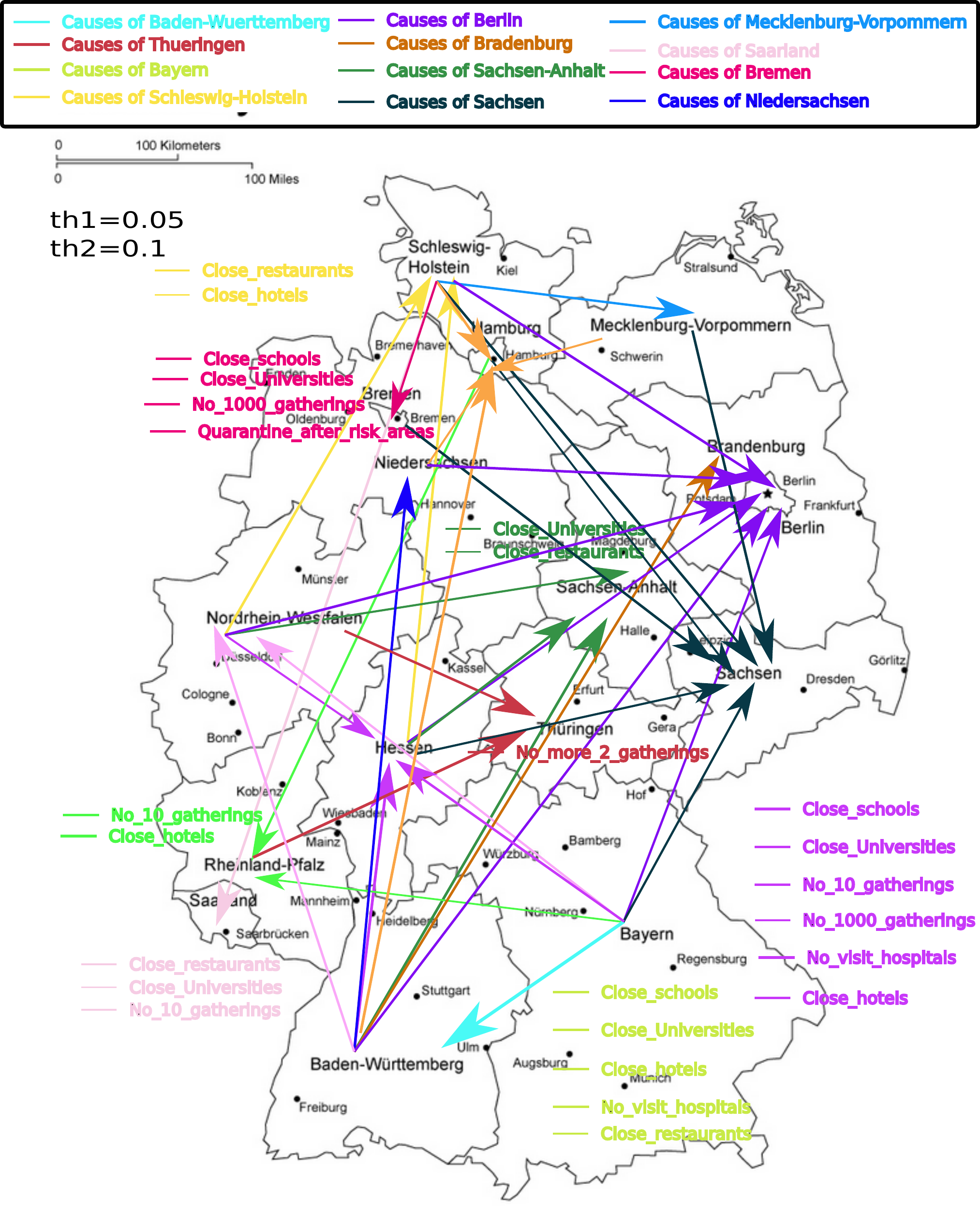} }}%
  \subfloat[District-level causal analysis.\label{districts}]{{\includegraphics[width=.5\textwidth]{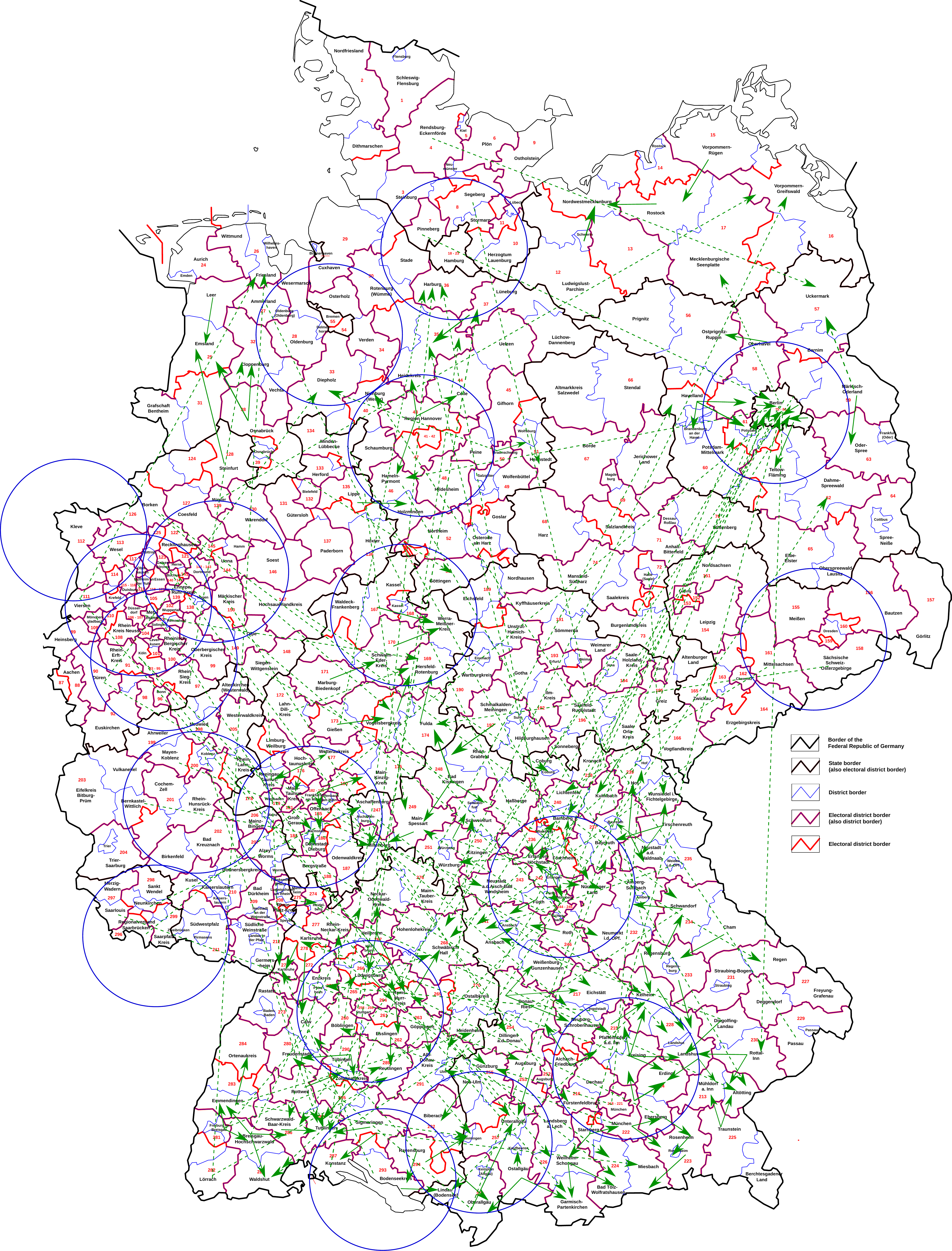} }}%
  \caption{a) Detected causal paths of the spread of Covid-19 among the German federal states, including causes among the restriction measures taken by each federal state. Each colour (in arrows and policies) indicates causes of one state (see top legend). These findings correspond to the looser of the four combinations of thresholds $(0.05, 0.1)$ that we tested. Results for the remaining three combinations can be found in Fig. \ref{federal_policies_all} in the supplement. b) Detected causal districts for the spread of Covid-19, for each district, using the modified SyPI algorithm. Solid arrows depict causes that are neighbour districts (i.e., sharing a common border). Dashed arrows depict causes that are not. The majority of the detected non-neighbour causes are close to cities with larger airports (MUC, STR, TXL, FDH, FMM, NUE, HAM, FRA, HHN, HAJ, NRN, CGN, DUC, DMT, DRS, BRE, KSF, SCN), and the majority of the detected causes are neighbours to the target. Note that since the dashed arrows are significantly longer than the solid ones, the Figure at first glance seems to show mostly dashed arrows. This is misleading; for a numeric comparison, see Figure \ref{categories}. Blue cycles indicate 40km radius around the largest airports. For the district-level analysis, the default thresholds of SyPI were used $(0.01, 0.2)$}%
\end{figure}

Table \ref{sypi_vs_granger_policies} in the supplement presents all the detected causes (states and policies) of each federal state, for the SyPI method and thresholds $(0.05, 0.1)$, as well as for the Lasso-Granger method.
With these "loose" thresholds we expect the largest number of detected causes with SyPI (corresponding to the bottom left map in Figure~\ref{foundedCausesWithPolicies}). We see that Lasso-Granger detects almost all candidates as causes. This indicates that this is a dataset with many latent confounders, which forces Granger to give incorrect causal claims. On the other hand, SyPI is robust against false positives due to latent common causes, and thus gives potentially more meaningful results.

\subsection{Enacted policies and causal roles of federal states}
Here we discuss the relation between the outcome of the above causal analysis and the applied restriction measures. This time, instead of looking for causes of Covid-19 cases in German federal states, we look at the states that helped contain the spread of the pandemic by not causing others. We make an observation that may serve as additional sanity check about the causal predictions of SyPI, using its stricter thresholds results:\footnote{Notice that this result should be treated with caution as it depends on the correctness of the above causal analysis and it may be confounded by the time order that the states reported causes.} states that were not found to cause other states were those that closed schools and universities ``early enough'' (meaning before $100$ cases were reported): Bremen, Th\"uringen, Saarland, Brandenburg, Sachsen-Anhalt and Sachsen (with the exception of Mecklenburg-Vorpommern). In addition, the German states that were found to cause others were also those that either did not take both measures combined (Schleswig-Holstein, Nordrhein-Westfalen, Rheinland-Pfalz \footnote{The arrow Rheinland-Pfalz $\rightarrow$ Th\"uringen does not appear in the subplot of strict thresholds because the p-value (0.011) for condition 1 was on the limit over the strict threshold 1 (0.01).}), or they took them relatively late (i.e., $> 100$ cases) (Bayern, Baden-W\"urttemberg, Hamburg, Hessen).

\subsection{Causal spread of Covid-19 among the German districts}\label{districts_explain_findings}
\begin{figure}
  \centering
  \subfloat[\label{categories}]{{\includegraphics[width=.4\textwidth]{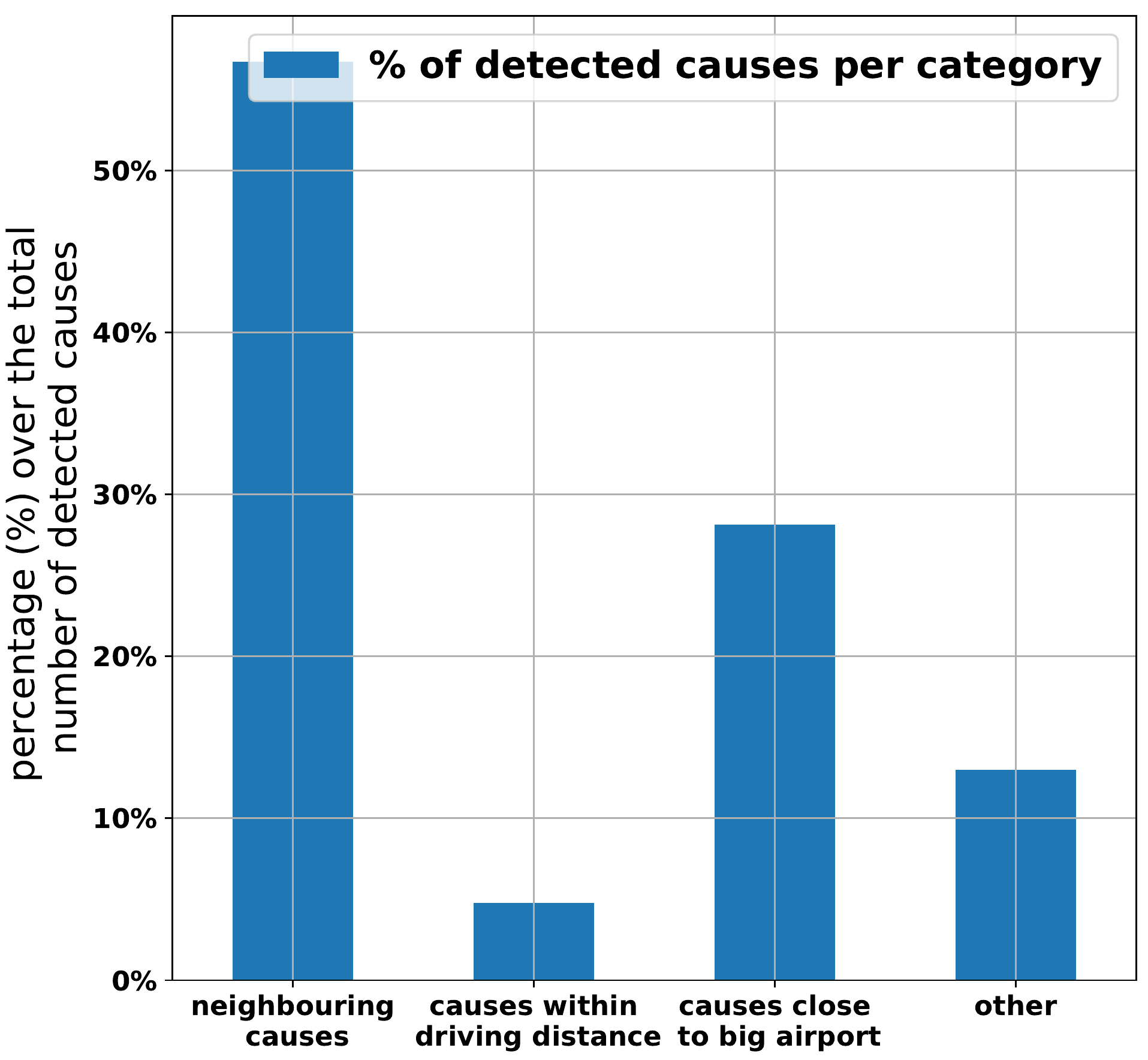} }}%
  \subfloat[\label{airports}]{{\includegraphics[width=.6\textwidth]{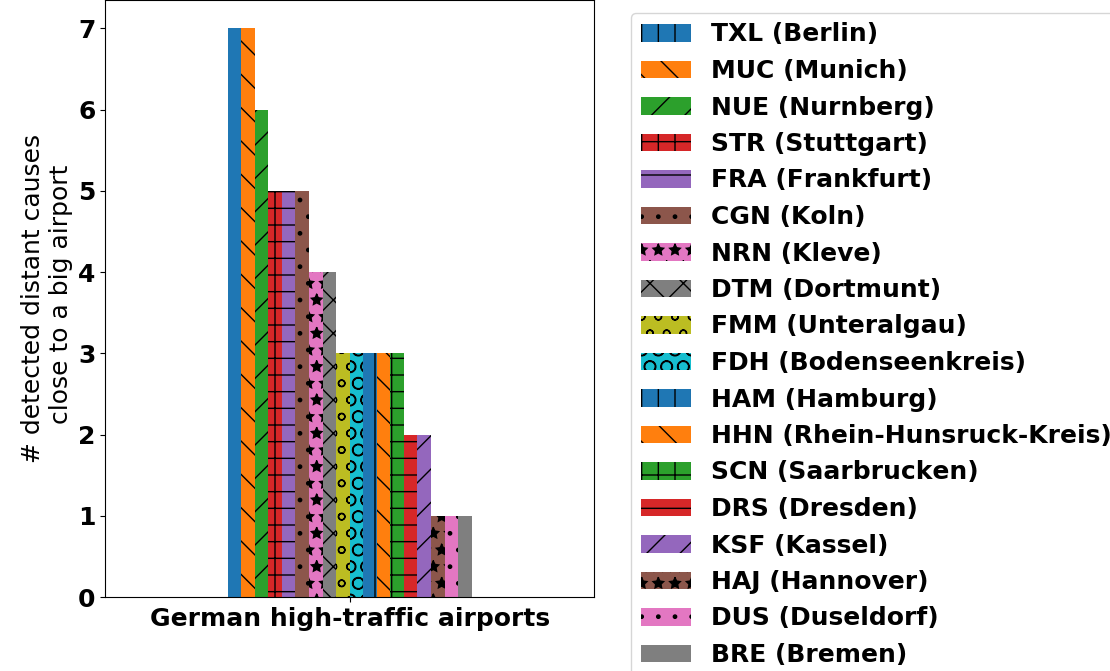} }}%
  \caption{a) Percentage of detected district causes (out of 231 causes detected in total) by category of district relative to target district. Most causes of a given target turn out to be neighbouring districts, and of the distant ones, many are close to major airports. Only 12\% of the detected causes cannot be justified by proximity to the target or to a larger airport. b) Detected distant causes located close to the large airports. To assign a detected causes to one of the airports it had to meet two criteria: 1.~located in near the airport (see main text), and 2.~the target of this cause also needs to be located close to another big airport. We sort the airports by the number of detected causes.}%
\end{figure}

Since the number of federal states is relatively small, we ran our analysis also at a finer level of granularity, using districts rather than states. 
Figure \ref{districts} depicts the map of Germany with all the detected causal districts for each district. Arrows with solid lines show neighbour causes, while arrows with dashed lines depict causes that do not share a border with the target. We see that for the majority of the target districts the detected causes are neighbouring districts, and that those that are not are generally near a large airport or within 40km distance from the target. Note that since the dashed arrows are significantly longer than the solid ones, the Figure at first glance seems to show mostly dashed arrows. This is misleading; for a numerical comparison, see Figure \ref{categories}. Distant causal districts often seem to be aligned with the routes of the domestic connections with the highest traffic, as reported in the DFS's latest flight report \cite{DFSreport}. These are: Berlin - Munich, Berlin - Frankfurt, D\"usseldorf - Munich, Cologne/ Bonn - Munich, D\"usseldorf - Berlin, Stuttgart - Hamburg, Frankfurt - Munich, Berlin - Stuttgart, with 1-2 million flights per year. These paths can be seen in Figure~\ref{districts}, made up of detected longer-range dashed causal arrows. Table \ref{sypi_landkreis} with the causal results shown in Figure \ref{districts} can be found in the supplement.

We categorise the total number of 231 causes detected into the following four categories: 1. Detected causes neighbouring (sharing common borders) the target district, 2. Detected causes near ($\sim$ 40km) to the target, 3. Detected causes close to a large airport, 4. Distant targets that cannot be categorised otherwise. As we can see in Figure~\ref{categories}, the majority of causes are neighbour districts, and only the 12\% of the causes cannot be justified by proximity to the target or a large airport. 
Figure~\ref{airports} shows the histogram of the detected causes that are located close to a large airport, in cases where also the target is reachable by another airport.

\section{Discussion}

\subsection{Findings of the causal analysis}\label{results_discussion}
We performed a causal analysis both on a federal state/policy level, and on a more fine-grained district level. We tried normalising the case numbers in different ways (e.g., dividing by the maximum), but the results were not much affected. We decided not to normalise the data by population, as we felt this would unduly enhance the influence of less populous states. 

For the policy analysis, we compared the findings of SyPI with the predictions of the widely used Lasso-Granger method \cite{Arnold2007}. We did not compare with seqICP \cite{pfister2019invariant}, which is a feature selection method, since this method requires that no interventions are applied on the target. In the present setting, the target always is subject to interventions. 
Lasso-Granger detected almost every candidate region and distancing measure as causal, which is not surprising in a confounded real-world dataset like the present one. 

\begin{figure}[H]
\centering
\includegraphics[width=0.85\linewidth]{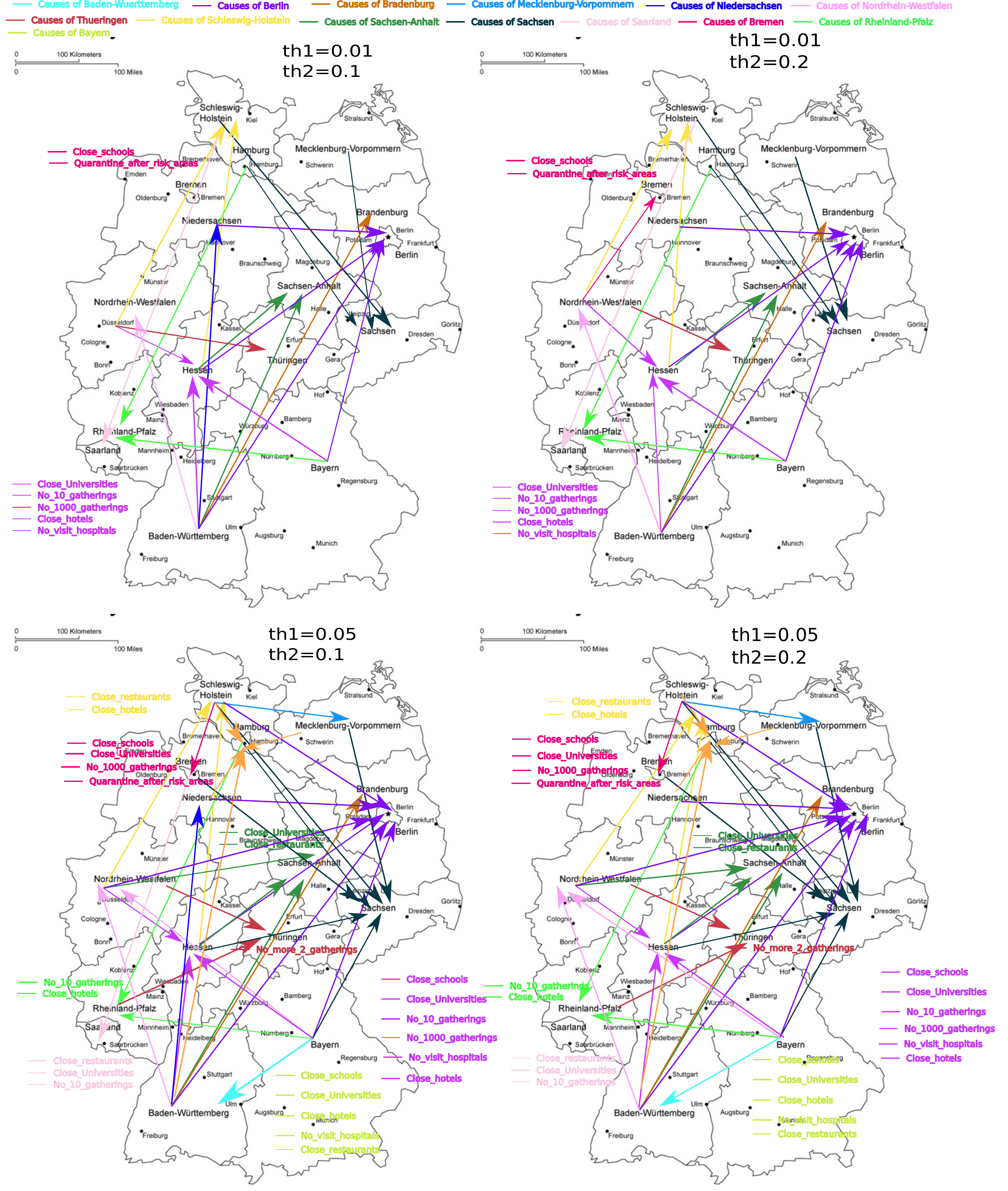}
\caption{Detected causal paths of the spread of Covid-19 among the federal German states, including causes among the restriction measures taken by each federal state. Each colour (in arrows and policies) indicates causes of one state (see top legend). The four subfigures correspond to the four combinations of threshold 1 and 2 that we tested. \label{federal_policies_all}}
\end{figure}

SyPI, on the other hand, yielded more meaningful results. We saw that the causes detected by SyPI on a district level tended to be neighbouring German districts, modulo the presence of major airports (which tend to be associated with industrial hubs). The pattern for federal states was consistent with this, but since there are fewer federal states than districts, numbers are small. The results in Figure~\ref{foundedCausesWithPolicies} seem meaningful in that much of the spread is local. In addition, Bayern and Baden-W\"urttemberg, the federal states with the largest current Covid-19 incidence,\footnote{\tiny{\url{https://www.rki.de/DE/Content/InfAZ/N/Neuartiges_Coronavirus/Situationsberichte/2020-05-31-en.pdf?__blob=publicationFile}}} have almost no arrows coming in from other states (see also Fig.~\ref{federal_policies_all}). In the district-level analysis only 38 out of 167 detected causes of targets in these two federal states belonged to another state. The majority of detected causes was due to internal mobility (84.7\% for Baden-W\"urttemberg and 73\% for Bayern). It is believed that those states (which lie in the South) had a strong influx of cases from Italy and Austria, where the pandemic took hold earlier.\footnote{\tiny{\url{https://www.rki.de/DE/Content/InfAZ/N/Neuartiges_Coronavirus/Situationsberichte/2020-03-13-en.pdf?__blob=publicationFile}}}
As a noteworthy detail, our algorithm identified Tirschenreuth (northeast Bavaria) as the cause of all its neighbouring districts (Wunsiedel im Fichtelgebirge, Bayreuth, Neustadt an der Waldnaab). On March 7th, a large beer festival took place in the town of Mitterteich in the district of Tirschenreuth, with a strong subsequent local COVID-19 outbreak.\footnote{\url{https://medicalxpress.com/news/2020-03-bavarian-town-germany-impose-full.html}}

Furthermore, we saw that for the federal states, different restriction policies were found as causal, yet the majority agreed on the importance of closing the universities and schools. 
Our findings about the causal role of banning gatherings of more than 1000 people, followed by closing of schools and ban of meetings of more than 2 people, are also in agreement with the modeling analysis of \cite{Dehningeabb9789}. 

\subsection{Validity of assumptions made}
A potential issue is the time delay between the application of a restriction measure and the observation of its effects on the target. Note that schools and universities were (often) closed later than some other measures were taken, e.g., the ban on larger gatherings. With SyPI, it is hard to infer which measure had the strongest effect unless we knew exactly the incubation time of each measure and actually shift the time series of cases by a corresponding amount. As we learn more about epidemiological parameters of Covid-19 (e.g., on the typical time delay between infection and being tested positive), we may be able to perform the latter analysis. 

As mentioned in Section \ref{assumptions}, we assume that the policies affect the target (Covid-19 cases) and not the other way around, in order to comply with our requirement that none of the descendants of the target belongs in its candidate causes. This may be violated if policies were adjusted based on the observed number of positive cases. We can be sure from the theoretical point of view that the detected causes are not confounded covariates. However, the method will likely have failed to detect all the true causes, if the aforementioned violation applies. With Theorem \ref{theorem3} we relaxed the strict assumption of SyPI about the target being a sink node, by requiring only that it has no descendants among its candidates. In practice, we try to ensure this by selecting as candidates only regions that have already reported cases before the target. This makes it likely that no (or few) effects of the target exist in its candidate causes.

\subsection{Contributions \& conclusions}
Motivated by an application on Covid-19 spreading, we stated and proved a Theorem relaxing the assumption of the causal feature selection algorithm SyPI of \cite{mastakouri2020necessary}, making it applicable to a causal analysis of daily reported Covid-19 cases of German states and districts and state-wise social distancing measures. While ground truth is not available, our results as discussed in Section~\ref{results_discussion} seem meaningful. Possibly the biggest weakness of our approach lies in the fact that the data we used is confined: (1) we only look at case numbers, in contrast to more sophisticated methods to track the spread of an epidemic using contact tracing or even genetic analyses \cite{neher_potential_2020}. Moreover, (2) the sample size is small (the pandemic still be relatively new), and (3) the political interventions considered are binary and thus also provide relatively little information. 
It is encouraging, however, that already such limited data seems to contain causal signals pertaining to a highly nontrivial task. This suggests that our approach may contribute towards meaningful causal analysis of political interventions on the spread of Covid-19 as more data becomes available.

The causal analysis proposed in this paper aims at contributing to the broader effort of scientists to understand the spread of the Covid-19 pandemic and the causal role of political interventions such as social distancing. The causal method being applied and assayed in this work can provide trustworthy causal results, since it is robust against false positives in the presence of latent confounders in time series --- note that in Covid-19 data science problems, with our limited present understanding, it is likely that relevant covariates are unobserved, leading to confounded problems. Despite the theoretical validity of the causal method, caution should be exercised in the interpretation of the results of the present study, due to the limited data available for this analysis (only daily reported Covid-19 cases for different regions, and some political interventions), and the sheer difficulty of the task. {\bf At present, we would thus not recommend that our empirical findings be used to guide public policy.} However, we find our results encouraging, given the hardness of causal structure learning from observational real-world data, known to practitioners in the field \cite{runge2019detecting}. We therefore believe that methods such as the one used above, and further developments based upon it, can contribute towards rational approaches for choosing and balancing restriction measures for pandemics such as Covid-19.

\section{Acknowledgements}
The authors would like to thank Alexander Ecker and his team. The political intervention data was collected upon  Prof.\ Ecker's initiative by Christina Th\"one and a team of volunteers at \url{http://crowdfightcovid19.org}. Thanks also go to Karin Bierig for help creating a database of neighbouring districts and airports, and to Dominik Janzing, Vincent Stimper, Simon Buchholz and Julius von K\"ugelgen for their helpful feedback on the manuscript.

\bibliography{myBib_nips2020}
\bibliographystyle{unsrtnat}
\urlstyle{same}
\section{Appendix/ Supplementary material for the paper: Causal analysis of Covid-19 spread in Germany}
\frenchspacing

\subsection{Results of causal analysis on federal level for all four combinations of thresholds for SyPI}

\newpage
\subsection{Theory}\label{theory}
\subsubsection{Definitions}
\begin{definition}[Causal Faithfulness]
A distribution $P$ is faithful to a directed acyclic graph (DAG) $G$ if no conditional independence relations other than the ones entailed by the Markov property are present.
\end{definition}\label{faithfulness}

\begin{definition}[Causal Markov Condition \cite{Spirtes1993}]
Let $G$ be a causal graph with vertex set $\mathcal{V}$ and $P$ be a probability distribution over the vertices in $V$ generated by the causal structure represented by $G$. $G$ and $P$ satisfy the Causal Markov Condition if and only if for every $W$ in $\mathcal{V}$, $W$ is independent of $\mathcal{V}\setminus(\text{Descendants}(W) \cup \text{Parents}(W))$ given Parents$(W)$.

Here we use the global version of the Markov condition, which reads: if $\mathcal{X} \!\perp\!\!\!\perp_{G} \mathcal{Y} \mid \mathcal{Z} \Rightarrow \mathcal{X} \!\perp\!\!\!\perp \mathcal{Y} \mid \mathcal{Z}$ for all disjoint vertex sets $\mathcal{X, Y, Z}$ (where $\!\perp\!\!\!\perp_{G}$ denotes d-separation, as defined above)\label{defCMC}
\end{definition}

\subsubsection{Proof of Theorem \ref{theorem3}}\label{proof}
\begin{proof}
The proof of Theorem 1 in \cite{mastakouri2020necessary} applies without changes.
Regarding Theorem 2 in \cite{mastakouri2020necessary}:
Assume that the direct path $X^i_{t} \rightarrow Y_{t+w_i}$ exists and it is unconfounded. Then, condition 1 of Theorem 2 in \cite{mastakouri2020necessary} is true. Now assume that condition 2 of Theorem 2 in \cite{mastakouri2020necessary} does not hold. 
This would mean that the set $\{\boldsymbol{S^i}, X^i_{t}, Y_{t+w_i-1}\}$ does not d-separate $X^i_{t-1}$ and $Y_{t+w_i}$. (Recall that a path $p$ is said to be \textit{d-separated} by a set of nodes in $Z$ if and only if $p$ contains a chain or a fork such that the middle node is in $Z$, or if $p$ contains a collider such that neither the middle node nor any of its descendants are in the $Z$.)
Hence, a violation of condition 2 would imply that (a) there is some middle node or descendant of a collider in $\{\boldsymbol{S^i}, X^i_{t}, Y_{t+w_i-1}\}$ and no non-collider node in this path belongs to this set, or (b) that there is a collider-free path between $X^i_{t-1}$ and $Y_{t+w_i}$ that does not contain any node in $\{\boldsymbol{S^i}, X^i_{t}, Y_{t+w_i-1}\}$. 
\begin{itemize}
\frenchspacing
\item[(a)] \textit{There is some middle node or descendant of a collider in $\{\boldsymbol{S^i}, X^i_{t}, Y_{t+w_i-1}\}$ and no non-collider node in this path belongs to this set:} the proof given in \cite{mastakouri2020necessary} remains unaffected if all $\mathbf{DE}_Y^{\mathcal{G}} \not \in \mathbf{X}$, because any collider $D$ or descendent of collider between some $X^j_{t}$ and $Y_{t+w_i}$ will be unobserved, therefore will not be possible to belong in the conditioning set $\{\boldsymbol{S^i}, X^i_{t}, Y_{t+w_i-1}\}$.
 
\item[(b)] \textit{There is a collider-free path between $X^i_{t-1}$ and $Y_{t+w_i}$ that does not contain any node in $\{\boldsymbol{S^i}, X^i_{t}, Y_{t+w_i-1}\}$:} the proof given in \cite{mastakouri2020necessary} remains unaffected.
\end{itemize}
\end{proof}

\subsection{Detailed findings from comparison of modified SyPI with Lasso-Granger of the Covid-19 spread among the German federal states.}
\begin{table}[H]
\centering
\caption{Detected causes for each federal German state (Bundesland), using SyPI with the loose combination of thresholds $(0.05, 0.1)$ (2nd column) and using Lasso-Granger (3rd column). As expected we see that the number of detected causes by Granger is multiple times more than those of SyPI; in most cases Granger detects as causes all the candidate states. Without knowing the ground truth, this is an obvious indication that the dataset includes hidden confounders, that make the federal states look all related to each other. This violation of causal sufficiency makes Granger to fail, as expected. On the other hand, SyPI does not suffer from such problems even when there are latent confounders.\label{sypi_vs_granger_policies}}
{\tiny\tabcolsep1.4mm%
\begin{filecontents*}{thr1_005_thr2_01_normalized.csv}
Target,Predicted causes by SyPI,Predicted causes by Granger
Bayern,{Close schools, Close universities, Close restaurants, Close hotels, Prohibit visits at hospitals},{Close schools, Close universities, No more than 1000 people gatherings, No more than 10 people gatherings, Quarantine 14 days after visiting risk areas, No more than 2 people gatherings, Close restaurants, Close hotels, Prohibit visits at hospitals}
Baden-Württemberg,Bayern,{Bayern, Close schools, Close universities, No more than 1000 people gatherings, No more than 10 people gatherings, Quarantine 14 days after visiting risk areas, No more than 2 people gatherings, Close restaurants, Close hotels, Prohibit visits at hospitals}
Nordrhein-Westfalen,{Bayern, Baden-Württemberg},{Bayern, Baden-Württemberg, Close schools, No more than 1000 people gatherings, No more than 10 people gatherings, Quarantine 14 days after visiting risk areas, No more than 2 people gatherings, Close restaurants, Close hotels, Prohibit visits at hospitals}
Hessen,{Bayern, Baden-Württemberg, Nordrhein-Westfalen, Close schools, Close universities, No more than 1000 people gatherings, No more than 10 people gatherings, Close hotels, Prohibit visits at hospitals},{Bayern, Baden-Württemberg, Nordrhein-Westfalen, Close schools, Close universities, No more than 1000 people gatherings, No more than 10 people gatherings, Quarantine 14 days after visiting risk areas, No more than 2 people gatherings, Close restaurants, Close hotels, Prohibit visits at hospitals}
Niedersachsen,Baden-Württemberg,{Bayern, Baden-Württemberg, Nordrhein-Westfalen, Hessen}
Schleswig-Holstein,{Nordrhein-Westfalen, Hessen, Close restaurants, Close hotels},{Bayern, Baden-Württemberg, Nordrhein-Westfalen, Hessen, Niedersachsen, Close schools, No more than 1000 people gatherings, No more than 2 people gatherings, Close restaurants, Close hotels, Prohibit visits at hospitals}
Berlin,{Bayern, Baden-Württemberg, Nordrhein-Westfalen, Hessen, Niedersachsen, Schleswig-Holstein},{Bayern, Baden-Württemberg, Nordrhein-Westfalen, Hessen, Niedersachsen, Schleswig-Holstein, Close schools, No more than 10 people gatherings, Close restaurants, Close hotels, Prohibit visits at hospitals}
Bremen,{Schleswig-Holstein, Close schools, Close universities, No more than 1000 people gatherings, Quarantine 14 days after visiting risk areas},{Bayern, Baden-Württemberg, Nordrhein-Westfalen, Hessen, Niedersachsen, Schleswig-Holstein, Berlin, Close schools, Close universities, No more than 1000 people gatherings, No more than 10 people gatherings, Quarantine 14 days after visiting risk areas, No more than 2 people gatherings, Close restaurants, Close hotels, Prohibit visits at hospitals}
Mecklenburg-Vorpommern,Schleswig-Holstein,{Bayern, Baden-Württemberg, Nordrhein-Westfalen, Hessen, Niedersachsen, Schleswig-Holstein, Berlin, Bremen, Close schools, Close universities, No more than 1000 people gatherings, No more than 10 people gatherings, Quarantine 14 days after visiting risk areas, Close restaurants, Close hotels, Prohibit visits at hospitals}
Hamburg,{Baden-Württemberg, Niedersachsen, Schleswig-Holstein, Mecklenburg-Vorpommern},{Bayern, Baden-Württemberg, Nordrhein-Westfalen, Hessen, Niedersachsen, Schleswig-Holstein, Berlin, Bremen, Mecklenburg-Vorpommern, Close schools, Close universities, No more than 1000 people gatherings, No more than 10 people gatherings, Quarantine 14 days after visiting risk areas, Close restaurants, Close hotels, Prohibit visits at hospitals}
Rheinland-Pfalz,{Bayern, Hamburg, No more than 10 people gatherings, Close hotels},{Bayern, Baden-Württemberg, Nordrhein-Westfalen, Hessen, Niedersachsen, Schleswig-Holstein, Berlin, Bremen, Mecklenburg-Vorpommern, Hamburg, Close schools, No more than 1000 people gatherings, No more than 10 people gatherings, Close restaurants, Close hotels, Prohibit visits at hospitals}
Sachsen,{Bayern, Hessen, Schleswig-Holstein, Bremen, Mecklenburg-Vorpommern, Hamburg},{Bayern, Baden-Württemberg, Nordrhein-Westfalen, Hessen, Niedersachsen, Schleswig-Holstein, Berlin, Bremen, Mecklenburg-Vorpommern, Hamburg, Rheinland-Pfalz, Close schools, Close universities, No more than 1000 people gatherings, Close hotels, Prohibit visits at hospitals}
Brandenburg,Baden-Württemberg,{Bayern, Baden-Württemberg, Nordrhein-Westfalen, Hessen, Niedersachsen, Schleswig-Holstein, Berlin, Bremen, Mecklenburg-Vorpommern, Hamburg, Rheinland-Pfalz, Sachsen, Close schools, Close universities, No more than 1000 people gatherings, No more than 10 people gatherings, Close restaurants, Close hotels, Prohibit visits at hospitals}
Saarland,{Schleswig-Holstein, Close universities, No more than 10 people gatherings, Close restaurants},{Bayern, Baden-Württemberg, Nordrhein-Westfalen, Hessen, Niedersachsen, Schleswig-Holstein, Berlin, Bremen, Mecklenburg-Vorpommern, Hamburg, Rheinland-Pfalz, Sachsen, Brandenburg, Close schools, Close universities, No more than 1000 people gatherings, No more than 10 people gatherings, Close restaurants, Close hotels, Prohibit visits at hospitals}
Sachsen-Anhalt,{Baden-Württemberg, Nordrhein-Westfalen, Hessen, Close universities, Close restaurants},{Bayern, Baden-Württemberg, Nordrhein-Westfalen, Hessen, Niedersachsen, Schleswig-Holstein, Berlin, Bremen, Mecklenburg-Vorpommern, Hamburg, Rheinland-Pfalz, Sachsen, Brandenburg, Saarland, Close schools, Close universities, No more than 1000 people gatherings, Quarantine 14 days after visiting risk areas, Close restaurants, Close hotels, Prohibit visits at hospitals}
Thüringen,{Nordrhein-Westfalen, Rheinland-Pfalz, No more than 2 people gatherings},{Bayern, Baden-Württemberg, Nordrhein-Westfalen, Hessen, Niedersachsen, Schleswig-Holstein, Berlin, Bremen, Mecklenburg-Vorpommern, Hamburg, Rheinland-Pfalz, Sachsen, Brandenburg, Saarland, Sachsen-Anhalt, Close schools, Close universities, No more than 1000 people gatherings, Quarantine 14 days after visiting risk areas, No more than 2 people gatherings, Close restaurants, Close hotels, Prohibit visits at hospitals}
\end{filecontents*}
\csvreader[
  tabular=|p{20mm}|p{40mm}|p{90mm}|,
  nohead,column count=3,
  table head=\hline,
  late after line=\\\hline,
  late after first line=\\\hline,
  table foot=\hline]{thr1_005_thr2_01_normalized.csv}%
{}%
{\csvlinetotablerow}
}

\end{table}

\subsection{Detailed findings from the causal analysis of the Covid-19 spread among the German district states.}
\begin{center}
\captionof{table}{Detected causes for each district German state, using SyPI. In the first column, the target district state is reported. In the second column, the detected causes among the neighbouring districts are reported. Finally, in the third column, we report the detected distant causes. Strict thresholds (the default of SyPI method) are used for the analysis. As explained in Section \ref{districts_explain_findings} and in Figure \ref{categories}, the majority of detected district causes are neighbours of the targets, and the majority of the distant detected causes are located close to a big airport.}\label{sypi_landkreis}

{\tiny\tabcolsep1.4mm%
\begin{filecontents*}{landkreise_sypiforrelese_20200128_20200515__p1_0_01__p2_0_2.csv}
Target Landkreis (German distric state),Causes found among neighbouring districts,Causes found among distant districts
SK Gelsenkirchen,[],[]
LK Landsberg a.Lech,[],[]
LK Starnberg,[],[]
LK Fürstenfeldbruck,[],SK Gelsenkirchen
SK München,[],[]
LK Traunstein,[],[]
SK Delmenhorst,[],[]
LK München,LK Starnberg,[]
LK Freising,LK München,[]
SK Köln,[],[]
LK Lippe,[],[]
LK Stormarn,[],[]
LK Ravensburg,[],[]
LK Göppingen,[],[]
LK Tübingen,[],[]
SK Freiburg i.Breisgau,[],[]
LK Rottweil,[],[]
LK Heinsberg,[],[]
LK Breisgau-Hochschwarzwald,SK Freiburg i.Breisgau,[]
LK Böblingen,[],LK München
SK Erlangen,[],[]
LK Ludwigsburg,[],SK Freiburg i.Breisgau
LK Viersen,[],[]
StadtRegion Aachen,[],[]
SK Kaiserslautern,[],[]
LK Wesel,[],[]
SK Hamburg,[],[]
LK Märkischer Kreis,[],[]
SK Fürth,[],[]
LK Heilbronn,LK Ludwigsburg,[]
LK Ostalbkreis,[],[]
LK Gießen,[],[]
SK Bonn,[],[]
LK Alb-Donau-Kreis,LK Göppingen,SK Fürth
LK Segeberg,[],[]
LK Rhein-Neckar-Kreis,[],LK Böblingen
SK Mönchengladbach,[],[]
LK Ostallgäu,[],[]
SK Lübeck,[],[]
SK Schwabach,[],[]
LK Lahn-Dill-Kreis,[],[]
SK Bremen,[],[]
SK Duisburg,[],[]
LK Oberhavel,[],[]
LK Düren,[],[]
LK Groß-Gerau,[],[]
SK Heilbronn,[],[]
SK Münster,[],[]
Region Hannover,[],[]
LK Borken,[],[]
SK Frankfurt am Main,[],[]
LK Herzogtum Lauenburg,[],[]
LK Hochtaunuskreis,[],[]
LK Zollernalbkreis,{LK Rottweil, LK Tübingen},[]
SK Nürnberg,SK Erlangen,{LK Segeberg, SK Gelsenkirchen}
LK Rheinisch-Bergischer Kreis,[],[]
SK Mannheim,[],[]
LK Rhein-Kreis Neuss,[],[]
LK Sächsische Schweiz-Osterzgebirge,[],[]
LK Ebersberg,LK München,[]
LK Cuxhaven,[],[]
LK Rosenheim,{LK Ebersberg, LK München},[]
SK Berlin Marzahn-Hellersdorf,[],[]
SK Berlin Mitte,[],[]
SK Berlin Neukölln,[],[]
SK Ulm,[],[]
LK Passau,[],[]
LK Saale-Orla-Kreis,[],[]
LK Lörrach,[],[]
LK Rems-Murr-Kreis,{LK Heilbronn, LK Ludwigsburg},{LK Ebersberg, LK Wesel}
LK Rhein-Sieg-Kreis,[],[]
LK Main-Kinzig-Kreis,SK Frankfurt am Main,[]
LK Pinneberg,[],[]
LK Esslingen,LK Rems-Murr-Kreis,[]
LK Bergstraße,[],[]
LK Karlsruhe,[],LK Freising
LK Oberbergischer Kreis,[],[]
LK Ammerland,[],[]
LK Vorpommern-Greifswald,[],[]
SK Bochum,[],[]
SK Berlin Tempelhof-Schöneberg,[],[]
LK Rotenburg (Wümme),[],[]
LK Mecklenburgische Seenplatte,[],SK Berlin Tempelhof-Schöneberg
LK Main-Tauber-Kreis,[],[]
LK Coesfeld,[],[]
SK Düsseldorf,[],[]
SK Berlin Pankow,[],SK Nürnberg
SK Stuttgart,[],[]
LK Emmendingen,{LK Breisgau-Hochschwarzwald, SK Freiburg i.Breisgau},[]
SK Berlin Friedrichshain-Kreuzberg,{SK Berlin Mitte, SK Berlin Tempelhof-Schöneberg},LK Starnberg
LK Sigmaringen,[],[]
LK Grafschaft Bentheim,[],[]
SK Mainz,[],[]
SK Heidelberg,SK Mannheim,[]
LK Bad Dürkheim,[],[]
LK Germersheim,[],[]
LK Neckar-Odenwald-Kreis,LK Heilbronn,LK Breisgau-Hochschwarzwald
LK Cham,[],[]
SK Koblenz,[],[]
SK Oldenburg,[],[]
LK Leer,[],[]
LK Aichach-Friedberg,[],[]
LK Vorpommern-Rügen,[],{LK Zollernalbkreis, SK Münster}
LK Roth,[],[]
LK Bodenseekreis,LK Ravensburg,[]
LK Osnabrück,[],[]
LK Stade,[],[]
LK Rhein-Erft-Kreis,[],[]
LK Rheingau-Taunus-Kreis,LK Hochtaunuskreis,[]
LK Neu-Ulm,SK Ulm,[]
LK Unna,[],[]
LK Weilheim-Schongau,LK Starnberg,LK Viersen
LK Waldeck-Frankenberg,[],[]
LK Oberallgäu,{LK Ravensburg, LK Ostallgäu},[]
LK Vogelsbergkreis,LK Gießen,LK Borken
LK Ortenaukreis,LK Emmendingen,[]
SK Berlin Reinickendorf,[],[]
LK Miesbach,LK Rosenheim,LK Sigmaringen
SK Braunschweig,[],[]
LK Dithmarschen,[],[]
LK Hohenlohekreis,[],[]
SK Dortmund,[],[]
LK Calw,LK Karlsruhe,LK Ravensburg
LK Bad Kissingen,[],[]
LK Euskirchen,[],[]
LK Celle,Region Hannover,[]
SK Würzburg,[],[]
LK Erlangen-Höchstadt,SK Erlangen,{LK Ammerland, SK Berlin Mitte}
LK Havelland,[],LK Ludwigsburg
LK Konstanz,LK Sigmaringen,[]
SK Ingolstadt,[],[]
LK Würzburg,[],LK Erlangen-Höchstadt
SK Karlsruhe,LK Karlsruhe,{LK Lahn-Dill-Kreis, LK Bodenseekreis}
SK Kempten,[],[]
SK Leipzig,[],[]
SK Augsburg,[],[]
LK Biberach,LK Neu-Ulm,[]
LK Minden-Lübbecke,[],[]
LK Bautzen,[],[]
LK Mettmann,[],[]
LK Harburg,SK Hamburg,{SK Erlangen, LK Gießen, LK Zollernalbkreis}
SK Berlin Charlottenburg-Wilmersdorf,[],[]
SK Bielefeld,[],[]
LK Herford,[],[]
LK Kassel,[],[]
SK Essen,[],[]
SK Rosenheim,LK Rosenheim,[]
SK Hof,[],[]
LK Warendorf,[],[]
SK Wilhelmshaven,[],[]
LK Rastatt,[],[]
LK Bitburg-Prüm,[],[]
LK Fürth,[],[]
LK Enzkreis,[],SK Ingolstadt
SK Dresden,[],[]
SK Baden-Baden,[],[]
LK Ennepe-Ruhr-Kreis,[],[]
LK Hildesheim,[],[]
LK Offenbach,SK Frankfurt am Main,[]
LK Steinfurt,[],[]
LK Schwarzwald-Baar-Kreis,LK Breisgau-Hochschwarzwald,{SK Berlin Reinickendorf, LK Rhein-Neckar-Kreis}
SK Erfurt,[],[]
LK Freudenstadt,LK Tübingen,LK Oberallgäu
LK Regensburg,LK Cham,LK Segeberg
LK Tuttlingen,{LK Schwarzwald-Baar-Kreis, LK Sigmaringen, LK Zollernalbkreis},{LK Germersheim, SK Koblenz, LK Vogelsbergkreis}
LK Pfaffenhofen a.d.Ilm,{LK Aichach-Friedberg, LK Freising},LK Bergstraße
LK Teltow-Fläming,SK Berlin Tempelhof-Schöneberg,[]
LK Schwandorf,LK Regensburg,LK Borken
LK Reutlingen,LK Esslingen,{LK Lörrach, SK Freiburg i.Breisgau}
LK Rostock,LK Vorpommern-Rügen,[]
LK Friesland,[],{LK Lörrach, LK Viersen}
SK Aschaffenburg,[],[]
SK Berlin Spandau,LK Havelland,LK Oberallgäu
LK Merzig-Wadern,[],[]
LK Spree-Neiße,[],[]
LK Saar-Pfalz-Kreis,[],[]
SK Osnabrück,[],[]
LK Schwäbisch Hall,LK Rems-Murr-Kreis,{LK Ludwigsburg, SK Hof, SK Berlin Friedrichshain-Kreuzberg}
LK Plön,[],[]
LK Dingolfing-Landau,[],[]
SK Offenbach,[],[]
LK Dachau,[],[]
LK Straubing-Bogen,[],[]
LK Saarlouis,[],[]
LK Stadtverband Saarbrücken,[],[]
LK Rottal-Inn,[],SK Berlin Charlottenburg-Wilmersdorf
SK Wiesbaden,[],[]
SK Bottrop,[],[]
LK Donau-Ries,LK Aichach-Friedberg,[]
LK Kelheim,{LK Freising, LK Pfaffenhofen a.d.Ilm, LK Regensburg},LK Düren
LK Landshut,{LK Dingolfing-Landau, LK Freising, LK Kelheim, LK Regensburg, LK Rottal-Inn},LK Vogelsbergkreis
SK Bremerhaven,[],[]
LK Leipzig,[],[]
SK Berlin Steglitz-Zehlendorf,[],{LK Stadtverband Saarbrücken, LK Pinneberg, LK Schwäbisch Hall}
LK Lindau,LK Bodenseekreis,LK Kelheim
LK Main-Spessart,LK Bad Kissingen,{SK Fürth, LK Lippe}
LK Marburg-Biedenkopf,[],[]
SK Berlin Lichtenberg,SK Berlin Marzahn-Hellersdorf,LK Fürth
SK Hagen,[],[]
LK Görlitz,[],[]
LK Garmisch-Partenkirchen,{LK Ostallgäu, LK Weilheim-Schongau},SK Heidelberg
LK Fulda,LK Bad Kissingen,LK Sächsische Schweiz-Osterzgebirge
LK Neunkirchen,[],[]
LK Mayen-Koblenz,[],[]
LK Neuwied,[],[]
LK Elbe-Elster,[],[]
LK Emsland,{LK Leer, LK Osnabrück, LK Steinfurt},[]
LK Oldenburg,[],[]
LK Neustadt a.d.Aisch-Bad Windsheim,[],LK Mayen-Koblenz
LK Erding,{LK Freising, LK Landshut, LK München},[]
LK Oberspreewald-Lausitz,[],[]
SK Pforzheim,[],[]
SK Berlin Treptow-Köpenick,SK Berlin Friedrichshain-Kreuzberg,[]
SK Krefeld,[],[]
LK Siegen-Wittgenstein,[],[]
SK Kiel,[],[]
LK Soest,[],[]
LK Westerwaldkreis,[],[]
SK Leverkusen,[],[]
SK Chemnitz,[],[]
SK Halle,[],[]
SK Weimar,[],[]
LK Waldshut,LK Breisgau-Hochschwarzwald,{SK Nürnberg, LK Schwandorf}
SK Weiden i.d.OPf.,[],[]
LK Tirschenreuth,[],[]
SK Solingen,[],[]
SK Rostock,[],[]
LK Vulkaneifel,[],[]
SK Frankenthal,[],[]
SK Magdeburg,[],[]
SK Remscheid,[],[]
LK Verden,[],[]
SK Eisenach,[],[]
LK Rhein-Hunsrück-Kreis,[],[]
LK Paderborn,[],[]
LK Burgenlandkreis,[],[]
LK Märkisch-Oderland,[],[]
LK Diepholz,[],{SK Braunschweig, LK Düren}
LK Forchheim,[],[]
LK Ostholstein,[],[]
LK Osterholz,[],[]
LK Oder-Spree,LK Märkisch-Oderland,[]
LK Hameln-Pyrmont,Region Hannover,SK Berlin Friedrichshain-Kreuzberg
LK Hochsauerlandkreis,[],[]
LK Ilm-Kreis,[],[]
LK Kitzingen,[],[]
LK Kleve,[],[]
LK Kyffhäuserkreis,[],[]
LK Main-Taunus-Kreis,LK Hochtaunuskreis,[]
LK Meißen,[],[]
LK Recklinghausen,[],[]
LK Bernkastel-Wittlich,[],[]
LK Neumarkt i.d.OPf.,[],LK Neustadt a.d.Aisch-Bad Windsheim
LK Bad Kreuznach,[],[]
LK Saale-Holzland-Kreis,[],[]
LK Aurich,[],[]
LK Salzlandkreis,[],[]
LK Amberg-Sulzbach,LK Schwandorf,[]
LK Saalekreis,[],[]
LK Barnim,[],[]
LK Bayreuth,LK Tirschenreuth,SK Augsburg
LK Mansfeld-Südharz,[],[]
LK Lüneburg,[],LK Konstanz
LK Anhalt-Bitterfeld,[],[]
SK Straubing,[],[]
LK Haßberge,[],[]
SK Wuppertal,[],[]
LK Kaiserslautern,[],[]
SK Schwerin,[],[]
LK Holzminden,LK Lippe,[]
LK Hof,LK Bayreuth,SK München
LK Aschaffenburg,[],LK Neunkirchen
SK Emden,[],[]
LK Mainz-Bingen,[],[]
SK Neustadt a.d.Weinstraße,[],[]
SK Gera,[],[]
SK Oberhausen,[],[]
LK Gifhorn,[],[]
SK Herne,[],[]
SK Salzgitter,[],[]
LK Augsburg,[],LK Saarlouis
SK Kassel,[],[]
SK Kaufbeuren,LK Ostallgäu,[]
LK Bad Tölz-Wolfratshausen,LK Weilheim-Schongau,{LK Bad Dürkheim, LK Landsberg a.Lech}
LK Deggendorf,[],[]
SK Ludwigshafen,[],[]
LK Cloppenburg,LK Osnabrück,LK Rhein-Sieg-Kreis
LK Börde,[],[]
LK Bamberg,{LK Erlangen-Höchstadt, LK Forchheim},SK Berlin Reinickendorf
SK Mülheim a.d.Ruhr,[],[]
LK Gütersloh,[],[]
LK Schweinfurt,{LK Bad Kissingen, LK Würzburg},SK Ulm
SK Cottbus,[],[]
LK Rendsburg-Eckernförde,[],[]
LK Northeim,[],[]
LK Wittmund,[],[]
LK Schleswig-Flensburg,[],[]
LK Uelzen,[],[]
LK Weißenburg-Gunzenhausen,LK Donau-Ries,[]
LK Nienburg (Weser),[],LK Passau
LK Unterallgäu,{LK Augsburg, LK Oberallgäu},SK Köln
LK Olpe,[],[]
LK Vechta,[],[]
LK Rhein-Lahn-Kreis,[],[]
LK Wetteraukreis,LK Hochtaunuskreis,[]
LK Regen,[],LK Siegen-Wittgenstein
LK Cochem-Zell,[],[]
LK Nordsachsen,[],[]
LK Hersfeld-Rotenburg,{LK Fulda, LK Vogelsbergkreis},{LK Lahn-Dill-Kreis, LK Mettmann}
LK Berchtesgadener Land,[],[]
LK Potsdam-Mittelmark,[],[]
LK Heidenheim,LK Emmendingen,[]
LK Ahrweiler,[],[]
LK Darmstadt-Dieburg,[],SK Köln
SK Landshut,[],[]
LK Südliche Weinstraße,[],[]
LK Nürnberger Land,LK Erlangen-Höchstadt,{LK Würzburg, LK Minden-Lübbecke}
LK Günzburg,LK Alb-Donau-Kreis,SK München
LK Göttingen,[],SK Osnabrück
LK Donnersbergkreis,[],[]
SK Hamm,[],[]
LK Freyung-Grafenau,[],[]
LK Dahme-Spreewald,[],SK Ulm
LK Harz,[],[]
LK Schwalm-Eder-Kreis,LK Marburg-Biedenkopf,[]
SK Passau,[],[]
LK Schmalkalden-Meiningen,[],[]
LK Altenkirchen,[],[]
SK Bamberg,[],[]
LK Altmarkkreis Salzwedel,[],[]
LK Alzey-Worms,[],[]
LK Miltenberg,{LK Aschaffenburg, LK Main-Spessart},LK Bernkastel-Wittlich
SK Trier,[],[]
LK Wittenberg,[],[]
LK Eichstätt,LK Donau-Ries,LK Aschaffenburg
LK Sankt Wendel,[],[]
LK Schaumburg,[],[]
LK Kusel,[],[]
LK Kulmbach,LK Hof,[]
LK Saalfeld-Rudolstadt,[],[]
LK Nordfriesland,[],[]
LK Rhön-Grabfeld,LK Schmalkalden-Meiningen,[]
LK Rhein-Pfalz-Kreis,[],[]
SK Regensburg,[],[]
LK Zwickau,[],[]
SK Suhl,[],[]
LK Peine,[],[]
SK Memmingen,LK Ravensburg,{LK Esslingen, LK Emmendingen}
LK Eichsfeld,[],[]
LK Steinburg,[],[]
SK Wolfsburg,[],[]
LK Altenburger Land,[],[]
SK Speyer,[],[]
SK Amberg,[],[]
LK Mittelsachsen,[],[]
LK Heidekreis,{LK Lüneburg, Region Hannover},{SK Nürnberg, LK Heilbronn}
SK Darmstadt,{LK Darmstadt-Dieburg, LK Offenbach},[]
LK Erzgebirgskreis,[],[]
LK Helmstedt,{SK Braunschweig, SK Wolfsburg},[]
LK Nordhausen,[],[]
LK Jerichower Land,[],[]
LK Kronach,LK Kulmbach,LK Berchtesgadener Land
LK Lichtenfels,[],LK Mittelsachsen
LK Limburg-Weilburg,[],[]
LK Goslar,[],[]
LK Ludwigslust-Parchim,[],[]
LK Neustadt a.d.Waldnaab,{LK Bayreuth, LK Tirschenreuth},[]
SK Worms,[],[]
LK Höxter,[],[]
LK Trier-Saarburg,[],[]
LK Neuburg-Schrobenhausen,LK Donau-Ries,SK Frankfurt am Main
SK Jena,[],[]
LK Coburg,[],[]
LK Gotha,[],[]
LK Greiz,[],[]
LK Odenwaldkreis,[],[]
LK Wartburgkreis,[],[]
SK Flensburg,[],[]
SK Landau i.d.Pfalz,[],[]
LK Vogtlandkreis,[],[]
LK Ansbach,{LK Fürth, LK Roth},{LK Dingolfing-Landau, SK Weiden i.d.OPf., LK Erzgebirgskreis}
SK Ansbach,LK Ansbach,[]
SK Brandenburg a.d.Havel,LK Havelland,[]
LK Wunsiedel i.Fichtelgebirge,LK Tirschenreuth,SK Halle
LK Unstrut-Hainich-Kreis,[],[]
LK Birkenfeld,[],[]
LK Weimarer Land,[],[]
LK Stendal,[],[]
SK Dessau-Roßlau,[],[]
LK Werra-Meißner-Kreis,{LK Kassel, LK Schwalm-Eder-Kreis},[]
SK Coburg,[],LK Hochsauerlandkreis
LK Nordwestmecklenburg,SK Schwerin,SK Rostock
LK Südwestpfalz,[],[]
SK Neumünster,[],[]
SK Potsdam,LK Havelland,SK Erlangen
LK Mühldorf a.Inn,{LK Landshut, LK Traunstein},[]
SK Schweinfurt,[],[]
SK Frankfurt (Oder),[],[]
LK Prignitz,[],[]
LK Altötting,{LK Rottal-Inn, LK Traunstein},{SK Heidelberg, LK Hameln-Pyrmont}
LK Wolfenbüttel,[],[]
LK Uckermark,[],{LK Märkisch-Oderland, LK Rendsburg-Eckernförde}
SK Bayreuth,[],LK Neustadt a.d.Waldnaab
LK Ostprignitz-Ruppin,LK Oberhavel,[]
LK Wesermarsch,[],[]
LK Dillingen a.d.Donau,[],LK Märkischer Kreis
SK Pirmasens,[],[]
LK Sömmerda,[],[]
LK Lüchow-Dannenberg,[],[]
LK Sonneberg,[],[]
LK Hildburghausen,[],[]
SK Zweibrücken,[],[]
\end{filecontents*}
\csvreader[
    longtable=|p{50mm}|p{50mm}|p{50mm}|,
    table head=
    \toprule\bfseries Target distric state &\bfseries Detected neighbouring causes & \bfseries Detected distant causes\\ 
    \midrule\endhead\bottomrule\endfoot,
    before reading={\catcode`\#=12},after reading={\catcode`\#=6}
    ]{landkreise_sypiforrelese_20200128_20200515__p1_0_01__p2_0_2.csv}{1=\ColOne, 2=\ColTwo, 3=\ColThree}
    {\ColOne & \ColTwo & \ColThree}
}

\end{center}

Here we provide figure \ref{districts} enlarged for better visibility. 
\begin{figure}[H]
\includegraphics[width=\textwidth]{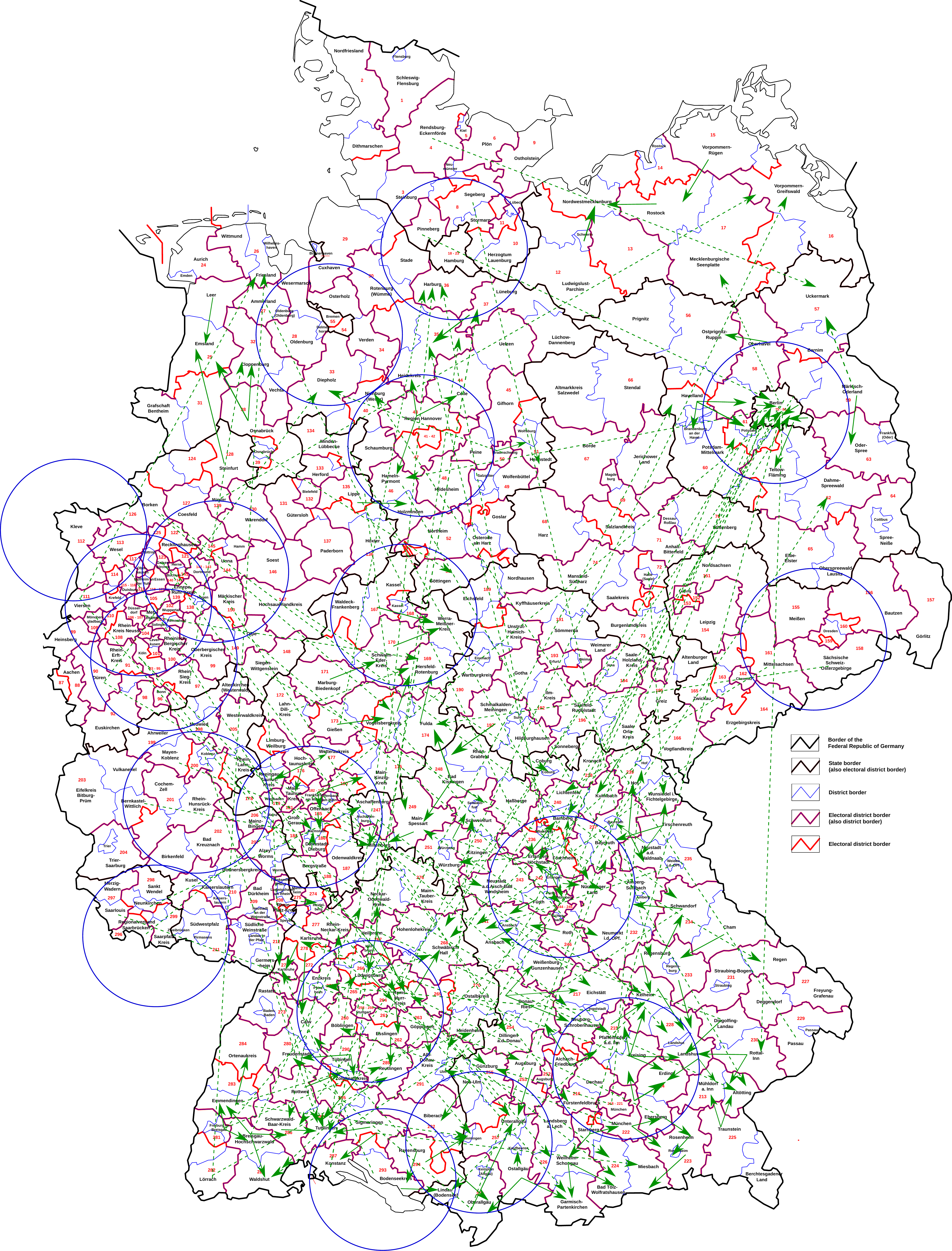} 
\caption{Detected causal districts for the spread of Covid-19, for each district, using the modified SyPI algorithm. Solid arrows depict causes that are neighbour districts (i.e., sharing a common border). Dashed arrows depict causes that are not. The majority of the detected non-neighbour causes are close to big cities with large airports (MUC, STR, TXL, FDH, FMM, NUE, HAM, FRA, HHN, HAJ, NRN, CGN, DUC, DMT, DRS, BRE, KSF, SCN), and the majority of the detected causes are neighbours to the target. Note that since the dashed arrows are significantly longer than the solid ones, the Figure at first glance seems to show mostly dashed arrows. This is misleading; for a numeric comparison, see Figure \ref{categories}. Blue cycles indicate 40km radius around the largest airports. For the district-level analysis, the default thresholds of SyPI were used $(0.01, 0.2)$}
\end{figure}

\end{document}